\documentclass[12pt]{article}
\usepackage{verbatim,color,amssymb,epsfig}
\usepackage[dvipsnames]{xcolor}
\usepackage{fancyhdr}
\usepackage{subcaption}
\usepackage[round, sort]{natbib}
\usepackage{algorithm}
\usepackage[normalem]{ulem}
\usepackage{amsmath}
\usepackage{threeparttable}
\usepackage{algpseudocode}
\usepackage{float}
\usepackage[colorlinks, linkcolor=red, citecolor=blue]{hyperref}
\usepackage{mathtools}
\usepackage{braket}
\usepackage{geometry}
\usepackage{subfloat}
\usepackage{booktabs}
\usepackage{longtable}
\usepackage{makecell}
\usepackage{authblk}
\usepackage{indentfirst}

\renewcommand{\baselinestretch}{1.5}

\def\boxit#1{\vbox{\hrule\hbox{\vrule\kern6pt
			\vbox{\kern6pt#1\kern6pt}\kern6pt\vrule}\hrule}}

\def\bse{\begin{eqnarray*}}
	\def\ese{\end{eqnarray*}}
\def\be{\begin{eqnarray}}
	\def\ee{\end{eqnarray}}
\def\bq{\begin{equation}}
	\def\eq{\end{equation}}
\def\bse{\begin{eqnarray*}}
	\def\ese{\end{eqnarray*}}

\def\b1e{{\mathbf e}}

\floatname{algorithm}{Algorithm}

\begin{document}

\def\spacingset#1{\renewcommand{\baselinestretch}%
{#1}\small\normalsize} \spacingset{1}

\title{\bf \Large Evaluating the Effectiveness of Public Policies on COVID-19 Containment: A PSM-DID Approach}
            \author{Zihan Wang}
            \affil{Department of Statistics and Data Science, Tsinghua University, Beijing 100084, China}

		\renewcommand\Affilfont{\itshape\small}
		\date{\vspace{-5ex}}
  \maketitle

  \bigskip
\begin{abstract}
The implementation of public policies is crucial in controlling the spread of COVID-19. However, the effectiveness of different policies can vary across different aspects of epidemic containment. Identifying the most effective policies is essential for providing informed recommendations for pandemic control. This paper examines the relationship between various public policy responses and their impact on COVID-19 containment. Using the propensity score matching-difference in differences (PSM-DID) model to address endogeneity, we analyze the causal significance of each policy on epidemic control.  Our analysis reveals that that policies related to vaccine delivery, debt relief, and the cancellation of public events are the most effective measures. These findings provide key insights for policymakers, highlighting the importance of focusing on specific, high-impact measures in managing public health crises.
\end{abstract}

\noindent%
\vfill

\newpage
\spacingset{1.7} 

\section{Introduction}
\label{sec1}
    The COVID-19 pandemic has profoundly impacted global life and economies since 2020, marking it as the most widespread and severe health crisis in the past century. Unlike the pandemics of the early 20th century, such as the influenza outbreak, the modern world faces new challenges due to increased global interconnectivity and changes in human lifestyles. These factors complicate the prevention and control of epidemics in our era. However, advancements in governance, biomedicine, and epidemiology have provided societies with more sophisticated tools to combat such crises. Countries worldwide have implemented a variety of public policies to curb the pandemic's spread, including lockdowns, medical interventions, and economic measures. The widespread rollout of vaccines, supported by biomedical advancements, has also become a crucial element in controlling the virus. Each nation’s public health strategy is shaped by factors such as population density, age demographics, ethnic diversity, and healthcare infrastructure, and these strategies often evolve as the pandemic progresses. The simultaneous implementation of multiple policies further complicates the analysis of any single policy’s effectiveness in controlling the epidemic.
    
    Numerous studies have examined the impact of public policies on COVID-19 containment using a variety of statistical models. For instance, \cite{page2021protective} employed propensity score matching to explore the relationship between state and local policies and COVID-19 mortality rates. \cite{sun2022quantifying} analyzed the effects of intervention timeliness, stringency, and duration on cumulative infections using counterfactual estimates. \cite{yang2021economic} applied the difference-in-differences method to assess how firm-level volatility responded to the COVID-19 shock through the lens of economic policy uncertainty. In another study, \cite{zhang2021effects} utilized structural equation modeling to provide scientifically grounded evidence for designing more effective COVID-19 policies in the transport and public health sectors. \cite{dzator2022policy} employed a panel data model to examine the effects of government policy stringency and handwashing measures.
    
    Other approaches include the use of negative binomial regression \citep{gaskin2021geographic}, latent factor models \citep{chen2022role}, and logistic regression \citep{liu2021covid} to study the impact of policies on daily life. Linear regression \citep{sylvester2021covid} and time series models \citep{pelagatti2021assessing} have been used to predict COVID-19 cases. Spatial modeling has also been employed; for example, \cite{guliyev2020determining} explored the spatial impact of various geographic factors on COVID-19, \cite{jiang2021spatial} provided insights on mobility restriction policies, and \cite{zhang2021generalized} used a generalized linear model for variable selection. Additionally, \cite{james2021use} developed an infectious disease model to investigate the effects of public policy interventions.
    
    In this paper, we explore the impact of public policies on the progression of the COVID-19 pandemic from multiple perspectives. By employing the widely used PSM-DID model in causal inference, we account for differences in national conditions across countries and analyze the effects of nearly 20 public policies, including lockdown measures, healthcare interventions, and economic policies, on the pandemic’s trajectory. Our goal is to identify which policies are effective and which may have limited impact. Overall, this study approaches the analysis of public policy effects on the pandemic through the lens of causal inference.
    
    The rest of the paper is organized as follows. Section \ref{sec.data} provides a detailed description of the data used in this study, including the specific definitions of each public policy variable, and presents visualizations of changes in the epidemic reproduction rate and virus variants across different countries. In Section \ref{sec.method}, we explain the application of the PSM-DID model to control for confounding factors between countries and assess the significance of each policy. Section \ref{sec.res} discusses the empirical results and their corresponding policy implications. Finally, Section \ref{sec.con} gives a brief conclusion of this study.

	\section{Data}
	\label{sec.data}
	In this section, we will give the sources, descriptive statistics and exploratory data analysis of the data used in the statistical modeling later. 
	
	\subsection{Data source and variable descriptions}
	To obtain the policy response intensity of governments in the past two years to deal with COVID-19, we use a popular public dataset called the Oxford COVID-19 Government Response Tracker (OxCGRT). The OxCGRT \citep{hale2021global} systematically collects information on several different common policy responses governments have taken, records these policies on a scale to reflect the extent of government action, and aggregates these scores into a suite of policy indices. Here, we choose 8 containment and closure policies (C), 2 economic policies (E), and 6 health system policies (H). Then, in order to measure the effect of policies, we use new confirmed cases of (7-day smoothed) per million people as the dependent variable, which is collected by the Center for Systems Science and Engineering (CSSE) at Johns Hopkins University. \cite{arroyo2021tracking} gave the real-time estimate of the effective reproduction rate (R) of COVID-19, which is used to detect the change points in Section \ref{subsec:dete}. In terms of country and date selection, we only retain the data of 38 European countries with not too small population and relatively complete data from March 15, 2020 to November 30, 2021. All the variable descriptions and basic statistics are shown in Table \ref{tab:var}. For detailed codebook, see \url{https://github.com/OxCGRT/covid-policy-tracker}. The GitHub repository, which includes the code and data for the empirical result reproduction of this paper, can be founded in \url{https://github.com/Wang-ZH-Stat/COVID-19}. 

	\begin{table}[H]
		\centering
		\small
            \caption{Variable descriptions  and basic statistics.}
			\label{tab:var}
		\begin{tabular}{lp{220pt}ccccc}
			\toprule
			Variable & Description & Mean & Sd. & Min. & Max. & \#$n$ \\
			\midrule
			C1 & Closings of schools and universities & 1.66 & 0.87 & 0 & 3 & 23729\\
			C2 & Closings of workplaces & 1.67 & 0.76 & 0 & 3 & 23732\\
			C3 & Canceling public events & 1.52 & 0.61 & 0 & 2 & 23732\\
			C4 & Limits on gatherings & 3.16 & 1.10 & 0 & 4 & 23732\\
			C5 & Closing of public transport & 0.43 & 0.59 & 0 & 2 & 23717\\
			C6 & Requirements of staying at home & 0.88 & 0.82 & 0 & 3 & 23707\\
			C7 & Restrictions on internal movement & 0.65 & 0.84 & 0 & 2 & 23662\\
			C8 & Restrictions on international travel & 2.61 & 0.87 & 0 & 4 & 23730\\
			E1 & If the government is providing direct cash payments to people who lose jobs & 1.55 & 0.66 & 0 & 2 & 23724\\
			E2 & If the government is freezing financial obligations for households & 1.24 & 0.76 & 0 & 2 & 23717\\
			H1 & Presence of public info campaigns & 1.96 & 0.21 & 0 & 2 & 23680\\
			H2 & Policies on who has access to testing  & 2.33 & 0.69 & 0 & 3 & 23709\\
			H3 & Policies on contact tracing after a positive diagnosis & 1.52 & 0.62 & 0 & 2 & 23625\\
			H6 & Policies on the use of facial coverings & 2.26 & 1.11 & 0 & 4 & 23690\\
			H7 & Policies for vaccine delivery & 2.00 & 2.10 & 0 & 5 & 23709\\
			H8 & Policies for protecting elderly people & 1.88 & 0.95 & 0 & 3 & 23693\\
			GRI & Overall government response index (all indicators) & 58.27 & 11.91 & 6.77 & 89.69 & 23740\\
			SI & Stringency index (all C indicators, plus H1) & 55.18 & 16.67 & 8.33 & 100.0 & 23742\\
			CHI & Containment and health index (all C and H indicators) & 57.54 & 11.99 & 7.74 & 90.00 & 23741\\
			ESI & Economic support index (all E indicators) & 63.46 & 27.26 & 0 & 100 & 23723\\
			NCSM & New confirmed cases of (7-day smoothed) per 1,000,000 people & 182.23 & 238.34 & 21.63 & 2078.77 & 23788\\
			R & Estimate of the effective reproduction rate & 1.08 & 0.31 & 0.09 & 3.69 & 23668\\
			\bottomrule
		\end{tabular}
	\end{table}

	\subsection{Exploratory data analysis}
\begin{figure}[H]
		\centering
		\includegraphics[width=1\textwidth]{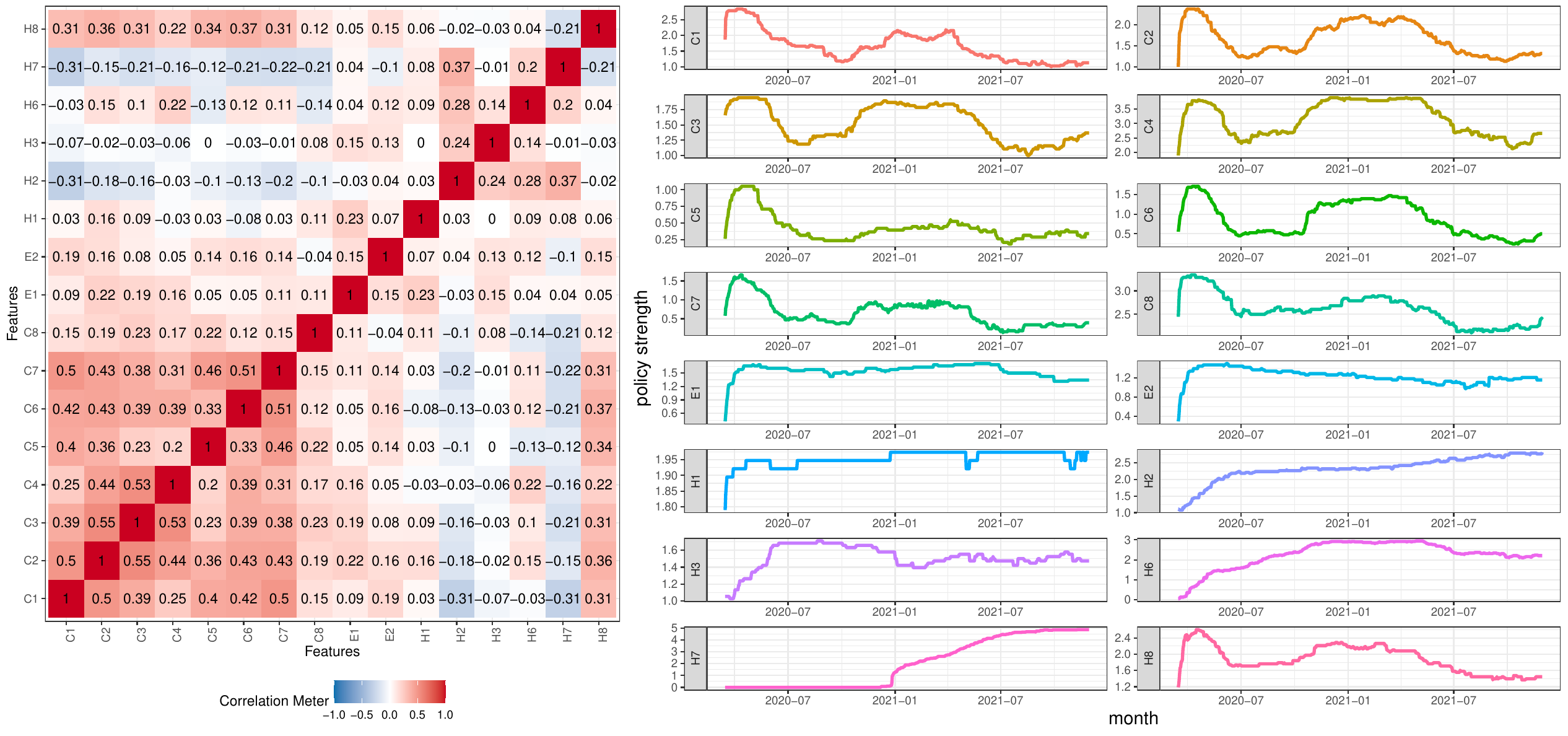}
		\caption{The correlation between policy responses and their respective trends over time. The data in the right panel averages 38 countries.}
		\label{fig:policy cor}
	\end{figure}

        In this section, we will show the preliminary results of visualizations of the data. In Figure \ref{fig:policy cor}, we illustrate the correlation between policy responses and their respective trends over time. We can find that there are always strong positive correlations within containment and closure policies, economic policies and health system policies, while there might be negatively correlated between containment and closure policies and health system policies. For containment and closure policies (denoted as ``C''), the policy intensity varies from top to bottom, which is determined by the degree of epidemic spread. Economic policies and health system policies basically remain at a relatively high level after May 2020.
	
\begin{figure}[H]
		\centering
		\includegraphics[width=1\textwidth]{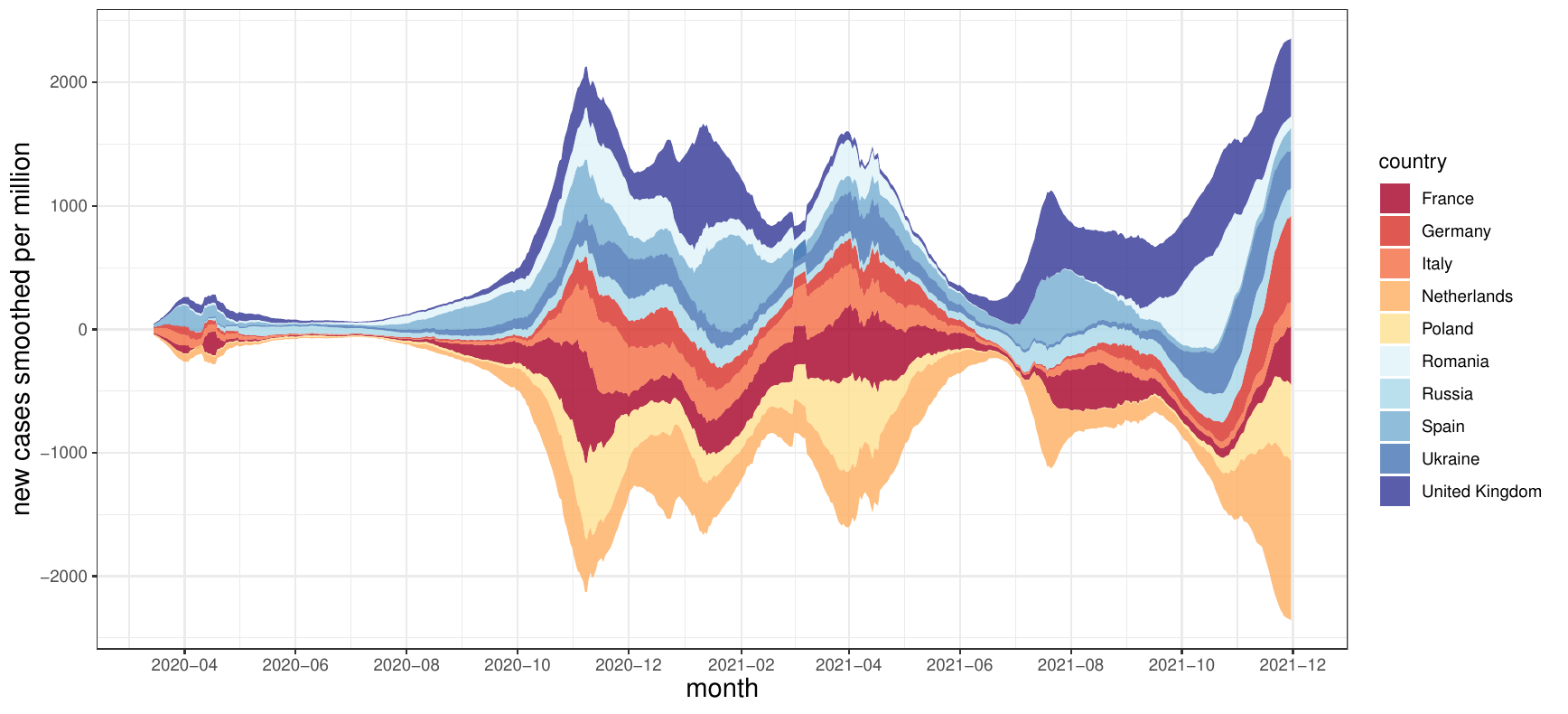}
		\caption{The steam plot of new cases smoothed per million in different countries. The new cases data are centralized for each date.}
		\label{fig:steam graph}
	\end{figure}

    As for the comparison at the country level, Figure~\ref{fig:steam graph} shows the new cases smoothed per million in different countries. Here we select the ten most populous countries, and daily data are centralized in these countries to clearly show the performance of these countries in epidemic containment. We can see that the UK, Romania and Spain played important roles in epidemic outbreak, while Netherlands and Poland performed relatively well.

\begin{figure}[H]
		\centering
		\includegraphics[width=1\textwidth]{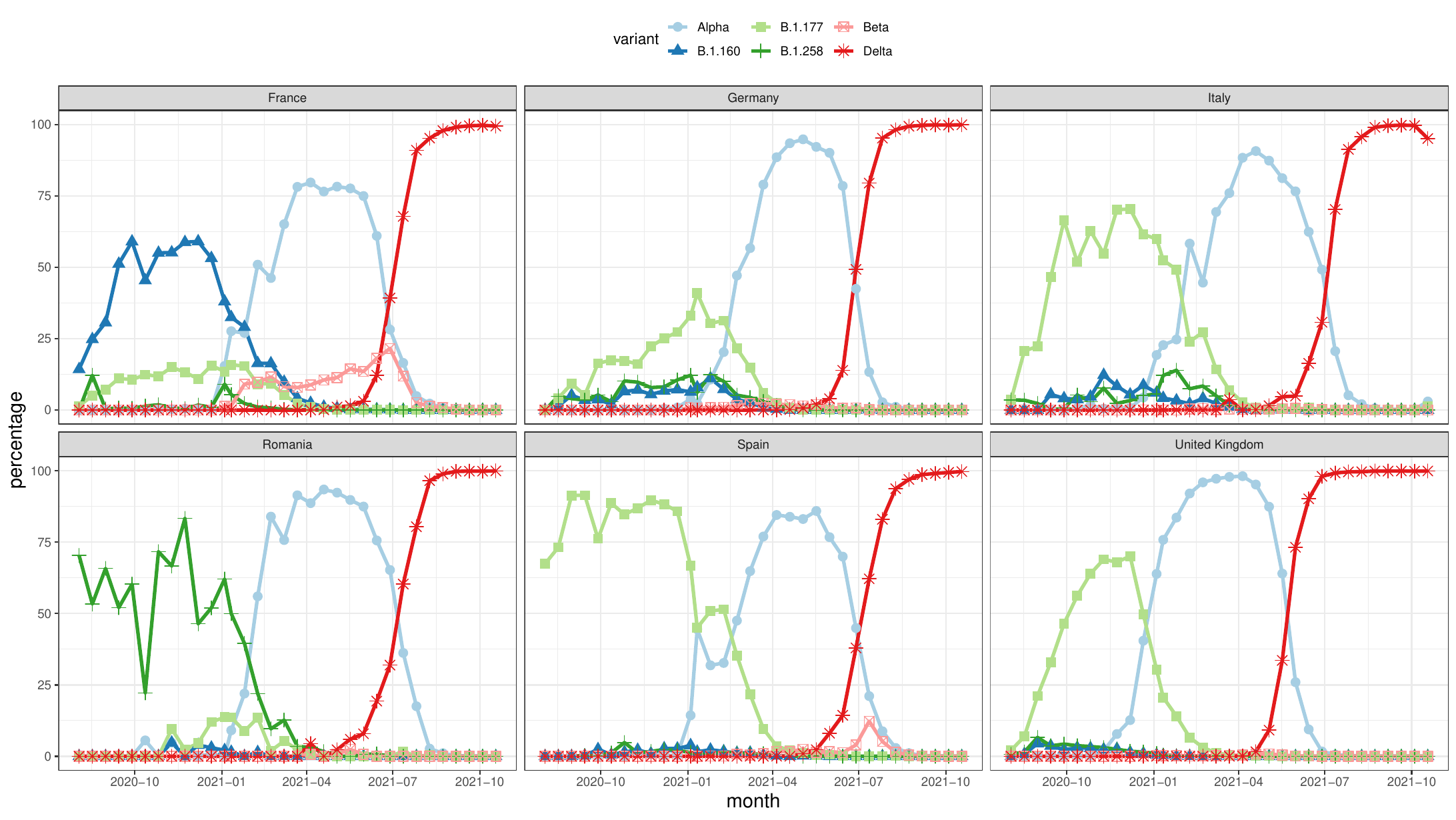}
		\caption{The trend of proportion of different variants over time in different countries. The data is collected from GISAID \citep{elbe2017data}.}
		\label{fig:variants}
	\end{figure}

        Figure \ref{fig:variants} gives the trend of proportion of different variants over time in six main countries. We can find that, before January 2021, the variants spread in these countries are different, including B.1.177, B.1.160 and B.1.258. From January 2021 to July 2021, the variant ``alpha'' originated from the UK accounted for a dominant proportion in all countries while since July 2021, the variant ``delta'' originated from India become the mainstream.
	
	\subsection{Detection of epidemic outbreak point}
	\label{subsec:dete}
\begin{figure}[H]
		\centering
		\includegraphics[width=1\textwidth]{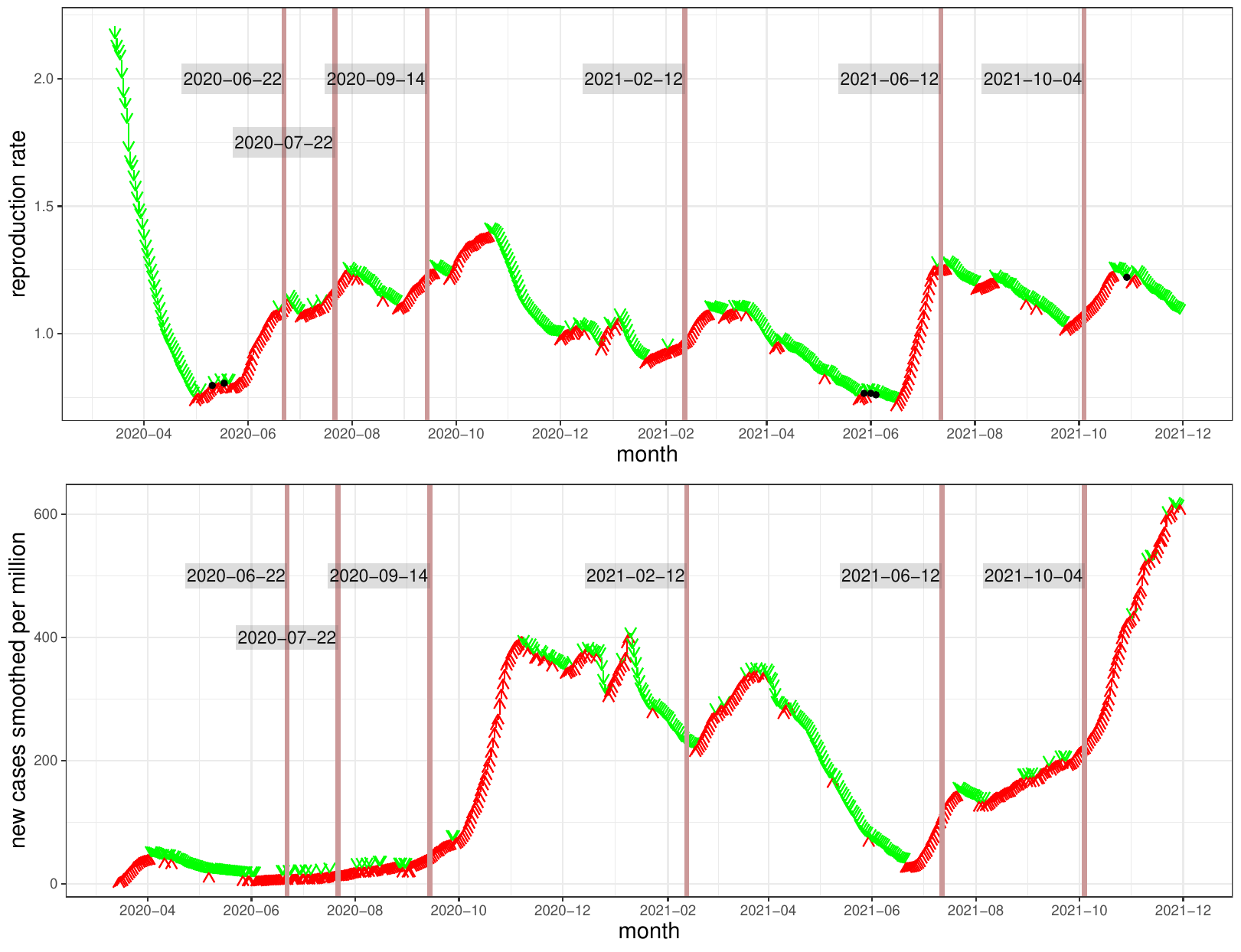}
		\caption{The results of sequential and multiple change point detection using the Mann-Whitney test statistic. Around the change points, the smoothed confirmed new cases present a piecewise linear trend.}
		\label{fig:change}
	\end{figure}

	In order to test the effect of policy response, we need to focus on a period of time before and after the epidemic outbreak. Here we consider the sequential and multiple change point detection for the effective reproduction rate. Since the reproduction rate data is obviously doesn't follow the normal distribution, we use the Mann-Whitney test statistic \citep{ross2011nonparametric} to detect location shifts in a stream with a (possibly unknown) non-Gaussian distribution. We only retain the 6 change points of effective reproduction rate in the rising stage, and the results are shown in Figure \ref{fig:change}. Then observing the position of these change points on the smoothed confirmed new cases curve, we find three main epidemic outbreak points, including ``2020-09-14'', ``2021-02-12'' and ``2021-10-04''. In the following sections, we mainly measure the effect the policy responses based on the epidemic date near these three points. Specifically, 30 days before and 30 days after the epidemic outbreak points will be studied. 
	
	\section{Methodology}
	\label{sec.method}
	In this section, we will introduce the most popular technique in empirical research to evaluate policy effects, which is propensity score matching-difference in differences (PSM-DID) model  \citep{hirano2003efficient}. This technique can be decomposed into two separated procedures. In this section, we illustrate the PSM procedure and DID procedure respectively, using the data of policy response C3 (canceling public events) and smoothed new confirmed cases near October 4, 2021 as an example. 
	
	\subsection{PSM procedure}
	\label{subsec:psm}
	Propensity score matching (PSM), which was introduced by \cite{rosenbaum1983central}, is a statistical matching technique that attempts to reduce the bias due to confounding variables. The possibility of bias arises because a difference in the treatment outcome between treated and untreated groups may be caused by a factor that predicts treatment rather than the treatment itself. For each covariate, randomization implies that treatment-groups will be balanced on average. Unfortunately, for observational studies, the assignment of treatments to research subjects is typically not random. Matching attempts to reduce the treatment assignment bias, and mimic randomization, by creating a sample of units that received the treatment that is comparable on all observed covariates to a sample of units that did not receive the treatment. 
	
	The covariates used to match countries are shown in Table \ref{tab:psm}. They are some macro factors that may affect the policy-making of government when the epidemic outbreaks.  

	\begin{table}
		\centering
		\small
            \caption{Macro covariates used to match countries.}
			\label{tab:psm}
		\begin{tabular}{lp{190pt}p{120pt}}
			\\
			
			\toprule
			Variable & Description & Source \\
			\midrule
			population & Population (latest available values) & United Nations\\
			population\_density & Number of people divided by land area & World Bank \\
			aged\_65\_older & Share of the population that is 65 years and older & World Bank\\
			gdp\_per\_capita & Gross domestic product at purchasing power parity & World Bank\\
			cardiovasc\_death\_rate & Death rate from cardiovascular disease in 2017 & Global Burden of Disease Collaborative Network\\
			diabetes\_prevalence & Diabetes prevalence (\% of population aged 20 to 79) in 2017 & World Bank\\
			hospital\_beds\_per\_thousand & Hospital beds per 1,000 people & OECD\\
			life\_expectancy & Life expectancy at birth in 2019 & United Nations\\
			human\_development\_index & A composite index measuring average achievement in three basic dimensions of human development & United Nations\\
			\bottomrule
		\end{tabular}
	\end{table}
	
\begin{figure}[H]
		\centering
		\subfloat[]{
			\begin{minipage}[H]{0.48\textwidth}
				\centering
				\includegraphics[width=1\textwidth]{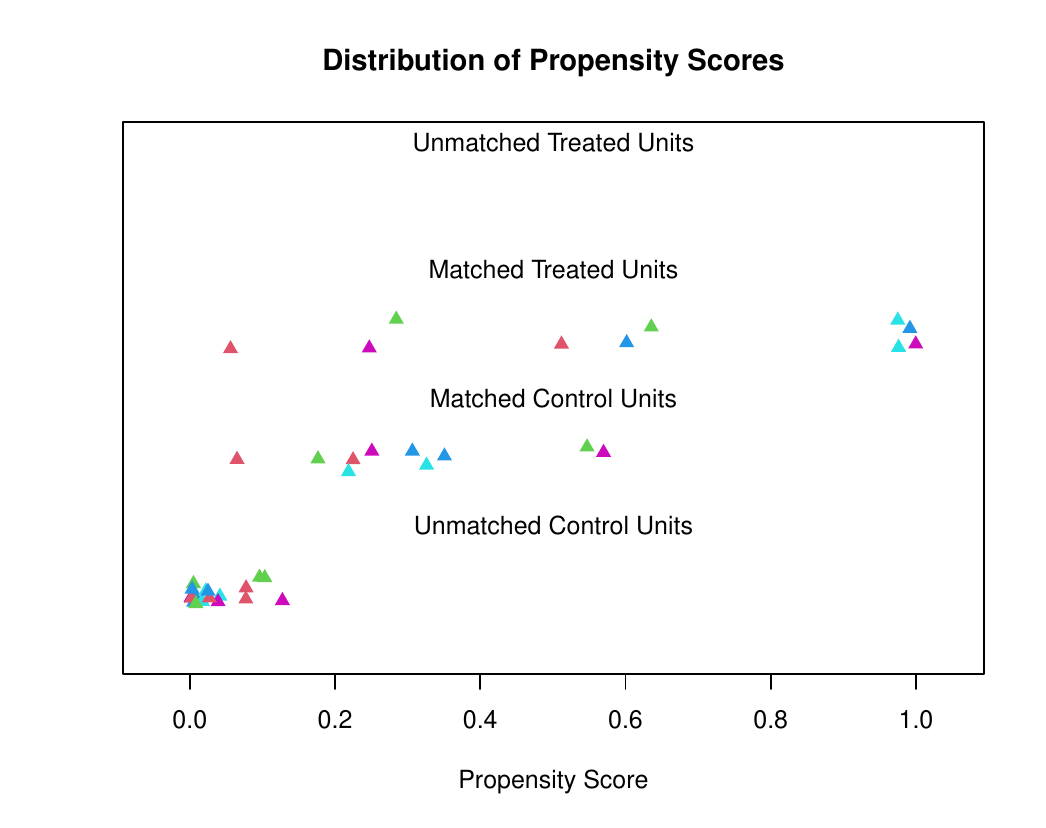}
			\end{minipage}%
		}%
		\subfloat[]{
			\begin{minipage}[H]{0.48\textwidth}
				\centering
				\includegraphics[width=1\textwidth]{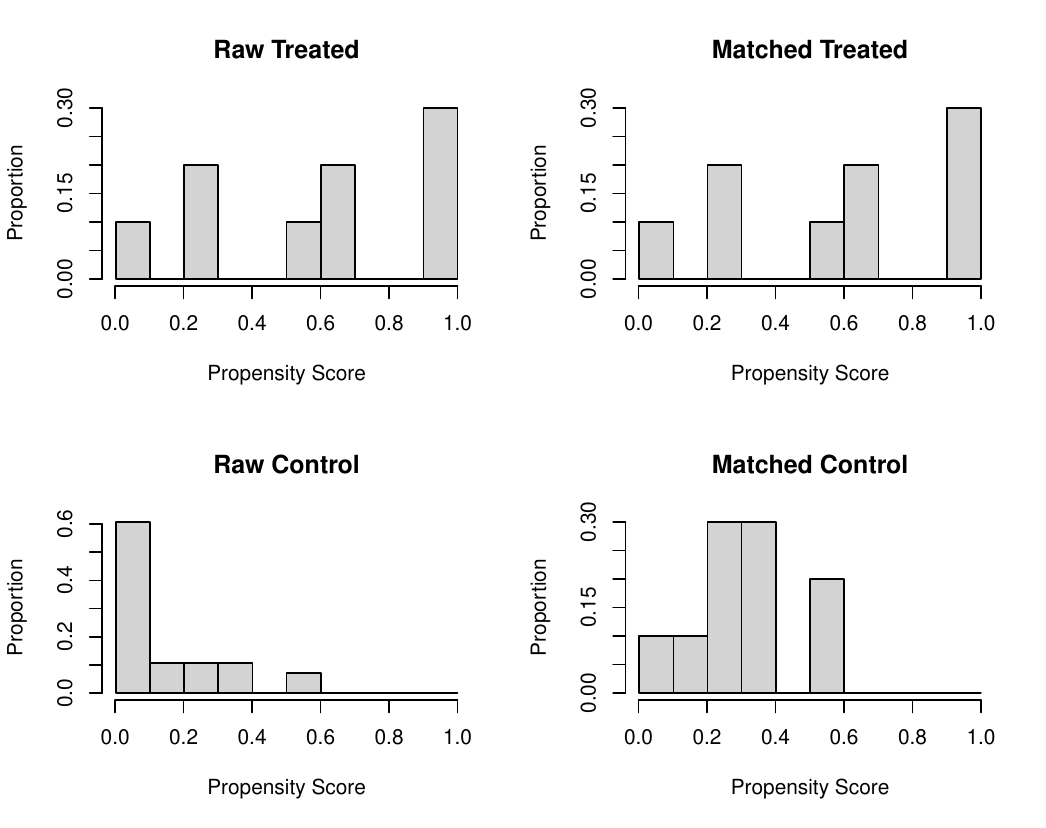}
			\end{minipage}%
		}%
		
		\subfloat[]{
			\begin{minipage}[H]{0.48\textwidth}
				\centering
				\includegraphics[width=1\textwidth]{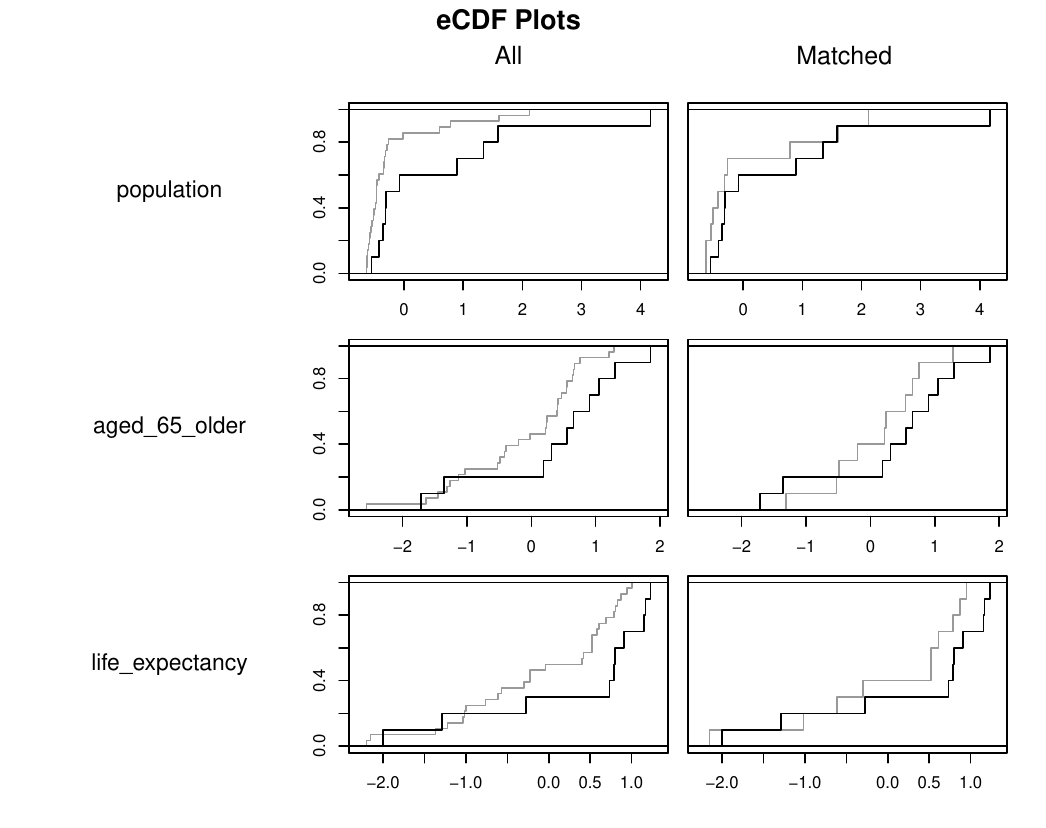}
			\end{minipage}
		}%
		\subfloat[]{
			\begin{minipage}[H]{0.48\textwidth}
				\centering
				\includegraphics[width=1\textwidth]{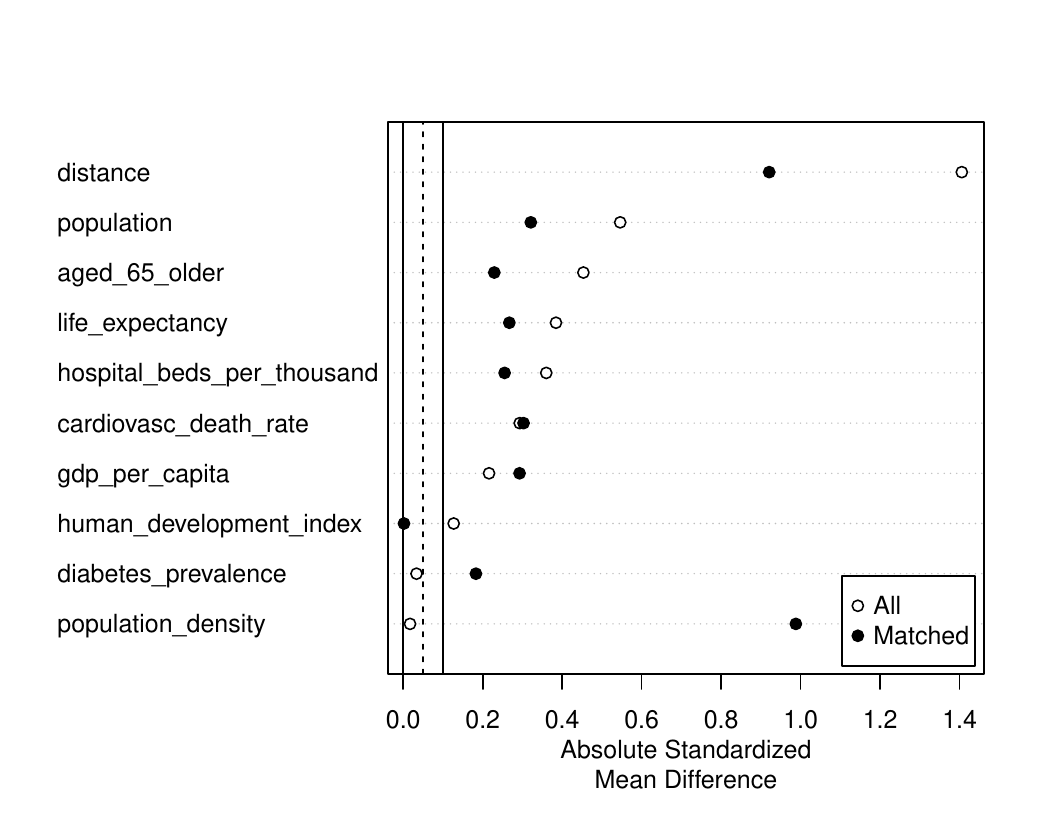}
			\end{minipage}
		}%
		
		\caption{The distribution of propensity scores, see subplots (a) and (b), and the comparison of covariates balance before and after matching, see subplots (c) and (d).}
		\label{fig:subfig} 
	\end{figure}

	In our illustration example (policy response C3, epidemic outbreak point ``2021-10-04''), 38 countries are divided into the treatment group and the control group according to whether the value of C3 at date ``2021-10-04'' is greater than 1, and we get 10 countries in the control group and 28 in the treatment group. We run a logistic regression and use the estimated probability of a given country coming from the treatment group as the propensity score. Then, we perform the optimal pair matching \citep{hansen2006optimal}. Each country in the control group has been matched to a similar country from the treatment group. The distribution of propensity scores and the differences of covariates between control group and treatment group before and after matching are shown in Figure \ref{fig:subfig}.

	\subsection{DID procedure}
	\label{subsec:did}
	Difference-in-difference (DID) evaluates the effect of a treatment on an outcome by comparing the average change over time in the outcome variable for the treatment group to the average change over time for the control group. In contrast to a time-series estimate of the treatment effect on units or a cross-section estimate of the treatment effect, DID uses panel data to measure the differences, between the treatment and control group, of the changes in the outcome variable that occur over time. DID is one of the four most commonly used methods for policy evaluation. The other three methods include matching, synthetic control and regression discontinuity, see \cite{imbens2009recent} for details. 
	
	There are many assumptions related to the DID model, and one of the most important is the parallel trend assumption. If individuals in the treatment group did not receive a treatment, which is counterfactual, the trends of response variable from treatment group and control group should be parallel. When this assumption is violated, we need to use PSM to reduce the endogeneity of the model. We only need to include the matched units to run a DID model. 
	
\begin{figure}[H]
		\centering
		\includegraphics[width=1\textwidth]{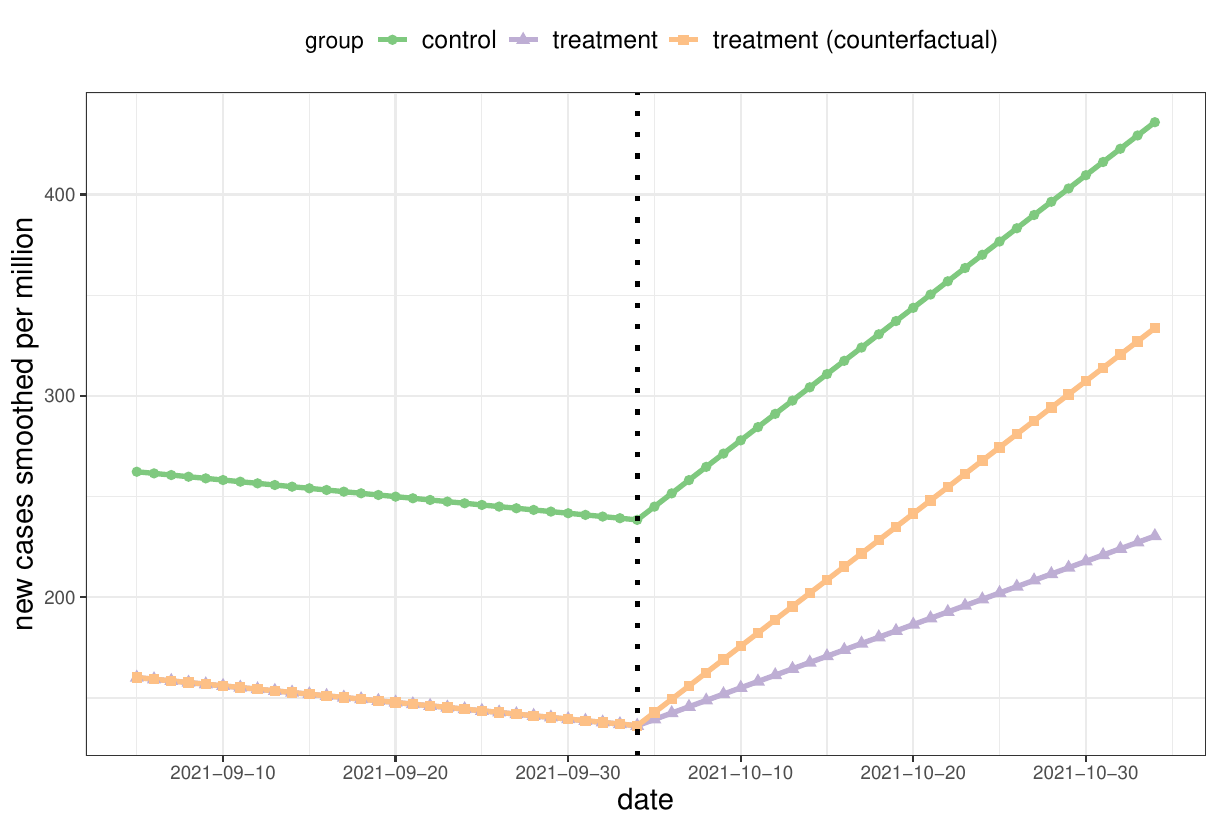}
		\caption{The fitted lines of the control group, the treatment group and the counterfactual treatment group.}
		\label{fig:did}
	\end{figure}

	In our illustration example, there are 10 units in both groups. By Figure \ref{fig:change}, the smoothed new confirmed cases seems to be piecewise linear before and after the outbreak epidemic point. For a given policy response, consider the piecewise linear regression
	\begin{equation}
		\label{eq:did}
		Y_{it}=\beta_0+\beta_1X_t+\beta_2P_{it}+\beta_3(X_t-X_{t_0})D_t+\beta_4(D_t-D_{t_0})D_tP_{it}+\varepsilon_t,
	\end{equation}
	where $Y_{it}$ is the smoothed new confirmed cases of country $i$, $X_t$ is the time variable taking the same values as $t$ from 1 to 60, $t_0$ is the policy response time (in our case, $t_0=30$), $P_{it}$ is a binary variable taking 0 if the country $i$ is in the control group and taking 1 if in the treatment group, $D_t$ is a dummy variable taking 0 when $1\le t\le 30$ and taking 1 when $31\le t\le60$, and $\varepsilon_t$ is the error term. Here, the negative significance of $\widehat{\beta}_4$ can be regarded as the evidence of significant policy response effect. To get obtain more informative measurement, we define the containment ratio (CR) as
	\begin{equation}
		{\rm CR}=\frac{\max(-\beta_4,0)}{\beta_1+\beta_3}\times 100\%,
	\end{equation}
	which illustrates the proportion of the slope of smoothed new confirmed cases after the outbreak point with time decreases by carrying out one given policy. The larger the value of CR, the more effective the policy is. In our example, $\widehat{\beta}_4=-3.4422$ with $p$-value taking 0.0041, and $\widehat{{\rm CR}}=52.31\%$. Canceling public events has a significant containment effect on the spread of the epidemic. The fitted lines of the control group, the treatment group and the counterfactual treatment group are shown in Figure \ref{fig:did}.
	
	\section{Results}
	\label{sec.res}

        In this section, we give the evaluation the effectiveness of most policy responses by PSM-DID technique. We use the similar procedures as those in Section~\ref{sec.method} on every policy response and three epidemic outbreak points. The results are shown in Table \ref{tab:didres}. Some policy response have been omitted, since we get three meaningless '/', implying that the intensity of this policy does not differ much among these European countries at the same date. We still divided 38 countries into control group and treatment group according to the intensity of policy response, and the division criteria are consistent at three time epidemic outbreak points. 
	\begin{table}[H]
		\centering
            \small
		\caption{The estimations of PSM-DID effects and containment ratios. * means $0.01\le p<0.05$, ** means $0.001\le p<0.01$ and *** $p<0.001$. In every cell of the table, the first row gives the estimation of $\beta_4$ in model (\ref{eq:did}), the second row gives the standard deviation, and the third row gives the estimation of containment ratio. '/' means that there are less than 3 countries in the control group or treatment group.}
		\label{tab:didres}
		\begin{tabular}{cccccccc}
			\toprule
			Policy & 2020-09-14 & 2021-02-12 & 2021-10-04 & Policy & 2020-09-14 & 2021-02-12 & 2021-10-04 \\
			\midrule
			C1 & \makecell[c]{0.1451 \\ (0.2426) \\ 0\%} & \makecell[c]{-0.4832 \\ (1.2224) \\ 11.07\%} & \makecell[c]{4.4789 \\ (1.6829) \\ 0\%} &
			C3 & \makecell[c]{-1.294*** \\ (0.2629) \\ 35.89\%} & \makecell[c]{-6.1438*** \\ (1.5246) \\ 100.99\%} & \makecell[c]{-3.4422** \\ (1.1956) \\ 52.31\%} \\

   \midrule
			C4 & \makecell[c]{-0.6398* \\ (0.2517) \\ 22.66\%} & / & \makecell[c]{-0.3187 \\ (1.0109) \\ 3.97\%} &
			C6 & / & \makecell[c]{-2.6323** \\ (0.9514) \\ 45.99\%} & \makecell[c]{-11.02** \\ (1.5776) \\ 93.11\%} \\

   \midrule
			E1 & \makecell[c]{1.7113*** \\ (0.2499) \\ 0\%} & \makecell[c]{0.2987 \\ (1.2112) \\ 0\%} & \makecell[c]{-0.0969 \\ (1.0466) \\ 1.18\%} &
			E2 & \makecell[c]{0.2083 \\ (0.2545) \\ 0\%} & \makecell[c]{-6.446*** \\ (1.0998) \\ 109.12\%} & \makecell[c]{-0.2798 \\ (1.1292) \\ 3.44\%} \\

   \midrule
			H2 & \makecell[c]{-1.6034*** \\ (0.2928) \\ 38.99\%} & \makecell[c]{-3.0598** \\ (1.1526) \\ 61.4\%} & \makecell[c]{0.2454 \\ (1.8826) \\ 0} &
			H7 & / & \makecell[c]{-2.3722* \\ (1.1726) \\ 48.6\%} & \makecell[c]{-5.5568*** \\ (1.4866) \\ 63.73\%} \\

   \midrule
			H8 & \makecell[c]{-0.3310 \\ (0.2709) \\ 10.75\%} & \makecell[c]{0.1783 \\ (1.3543) \\ 0\%} & \makecell[c]{-4.0268*** \\ (1.1184) \\ 36.5\%} & & & & \\
			
			\bottomrule
		\end{tabular}
	\end{table}
	
	For containment and closure policies, closings of schools and universities (C1) and limits on gatherings (C4) have poor effectiveness, and canceling public events (C3) and requirements of staying at home (C6) are significantly effect. One possible explanation can be that canceling public events and requirements of staying at home are more operable and supervised than limits on gatherings, and closings of schools and universities has little effect on containing the epidemic. 
	
	For economic policies, debt and contract relief (E2) is more effective than income support (E1). Even E1 is significantly positive at ``2020-09-14''. One possible explanation can be that providing more cash support to idle people will encourage them to gather for fun, while debt and contract relief can ensure that people are not forced to go out to make a living. 
	
	For health system policies, testing policy (H2) is much effective in the first two epidemic outbreak points, and protection of elderly people (H8) becomes more important in recent days. Vaccination policy (H7) is always effect, which is reasonable. With the normalization of the epidemic, the attitudes of people towards testing has become more passive. Perhaps in the future, the government can pay more attention to the protection of elderly people and carry out more refined policies. 
	
	Based on the containment ratios, the four most effective policy responses are canceling public events, requirements of staying at home, testing policy and vaccination policy.

	\section{Conclusions}
	\label{sec.con}

        In this paper, we examine the impact of public policies on epidemic containment using causal inference techniques. The PSM-DID model, which mimics randomization, is employed to assess the effectiveness of various policy responses. Our analysis reveals that for containment measures, canceling public events and enforcing stay-at-home requirements are more effective than limiting gatherings. In terms of economic policies, debt relief proves to be more impactful than direct cash support. For health system policies, combining vaccination efforts with protections for the elderly can lead to more effective outcomes.

        Despite the insights provided by the causal inference method, some limitations exist. First, the PSM-DID model assumes that units are uncorrelated, but there may be spatial dependencies among European countries that warrant consideration of spatial models. Second, since countries often implement multiple policies simultaneously, analyzing each policy in isolation may not fully capture their combined effects. Integrated policy indices, such as the GRI, could offer a more comprehensive alternative. Lastly, studying multiple countries and policies concurrently presents challenges due to numerous confounding factors. The synthetic control method proposed by \cite{abadie2003economic} could be a useful approach for evaluating the effects of multiple policies within a single country.

	\linespread{1}\selectfont
	\bibliographystyle{apalike}
	\bibliography{ref}

\end{document}